\documentclass[aps,pra,floatfix,reprint, superscriptaddress]{revtex4-2}
\usepackage{graphicx}
\usepackage{natbib}
\usepackage{braket}
\usepackage{amsmath} 
\usepackage{amssymb}
\usepackage{bm}
\usepackage[separate-uncertainty=false, multi-part-units = single, product-units=single, table-align-exponent]{siunitx}
\usepackage{upgreek}
\usepackage{color, soul}
\usepackage{dcolumn}
\newcolumntype{d}[1]{D..{#1}}

\begin{document}

\title{Precision measurement of the \textsuperscript{43}Ca\textsuperscript{+} nuclear magnetic moment }

\author{R. K. Hanley}
\affiliation{Department of Physics, University of Oxford, Clarendon Laboratory, Parks Road, Oxford OX1 3PU, U.K.}
\author{D. T. C. Allcock}
\affiliation{Department of Physics, University of Oxford, Clarendon Laboratory, Parks Road, Oxford OX1 3PU, U.K.}
\affiliation{Department of Physics, University of Oregon, Eugene, OR, USA}
\author{T. P. Harty}
\affiliation{Department of Physics, University of Oxford, Clarendon Laboratory, Parks Road, Oxford OX1 3PU, U.K.}
\author{M. A. Sepiol}
\affiliation{Department of Physics, University of Oxford, Clarendon Laboratory, Parks Road, Oxford OX1 3PU, U.K.}
\author{D. M. Lucas}
\email{david.lucas@physics.ox.ac.uk}
\affiliation{Department of Physics, University of Oxford, Clarendon Laboratory, Parks Road, Oxford OX1 3PU, U.K.}

\date{\today}

\begin{abstract}
We report precision measurements of the nuclear magnetic moment of \textsuperscript{43}Ca\textsuperscript{+}, made by microwave spectroscopy of the 4s $^2$S$_{1/2}$ $\left|F=4, M=0\right\rangle \rightarrow \left|F=3, M=1\right\rangle$ ground level hyperfine clock transition at a magnetic field of $\approx \SI{146}{G}$, using a single laser-cooled ion in a Paul trap. We measure a clock transition frequency of $f = \SI{3199941076.920 \pm 0.046}{Hz}$, from which we determine $\mu_I / \mu_{\rm{N}} = -1.315350(9)(1)$, where the uncertainty (9) arises from uncertainty in the hyperfine $A$ constant, and the (1) arises from the uncertainty in our measurement. This measurement is not corrected for diamagnetic shielding due to the bound electrons. We make a second measurement which is less precise but agrees with the first. We use our $\mu_I$ value, in combination with previous NMR results, to extract the change in shielding constant of calcium ions due to solvation in D$_2$O: $\Delta \sigma = -0.00022(1)$.

\end{abstract}

\maketitle
The favourable scaling laws of bound-state QED (BSQED) effects with proton number $Z$ have made highly-charged hydrogen- and lithium-like ions ideal probes of fundamental theories of atomic constituents, leading to significant theoretical \cite{Shabaev2015} and experimental \cite{Schabinger2012, Wagner2013, Sturm2013, Sturm2017} attention. An overview of tests of BSQED are discussed in detail in recent reviews by Kozlov \textit{et al.} \cite{Kozlov2018} and Indelicato \cite{Indelicato2019}. However, as with the proton-size puzzle preventing further improvements in tests of QED \cite{Pohl2013, Carlson2015, Hammer2020}, progress in stringent tests of BSQED is limited by our understanding of finite nuclear-size effects \cite{Gustavsson1998}. The dominant source of error \cite{Shabaev1997} is the Bohr-Weisskopf effect \cite{Bohr1950, Buttgenbach1984}, which describes the spatial distribution of the nuclear magnetisation. Typically one relies upon measured nuclear magnetic moments $\left(\mu_I\right)$ to infer the theoretical Bohr-Weisskopf correction. The precision at which one can test BSQED thus relies upon the precision of the known $\mu_I$. The importance of such measurements is exemplified by the `Bismuth Hyperfine Puzzle', where an incorrect measurement of $\mu_I$ in $^{209}$Bi resulted in a 7$\sigma$ difference between experimental and theoretical predictions \cite{Gustavsson1998, Ullmann2017, Skripnikov2018}.

In addition to tests of BSQED, highly-charged ions have also been used to probe nuclear structure \cite{Brown2001, Caurier2005, Otsuka2020}. Of particular interest are ions with magic numbers of nucleons \cite{Mayer1949}. Recently there has been significant theoretical \cite{Rutz1998, Sahoo2009, Holt2012, Hagen2012, Roth2012, Holt2014, Soma2014} and experimental attention \cite{Speidel2003, GarciaRuiz2015, Kohler2016, Ruiz2016, Klose2019} given to the calcium isotopic chain as there exist two naturally-occurring doubly-magic isotopes, $^{40}$Ca and $^{48}$Ca. The properties of the calcium isotopic chain can reveal new aspects of nuclear forces, such as three-body contributions \cite{Hammer2013} and the appearance of new magic numbers at extreme neutron-to-proton ratios \cite{Ozawa2000, Steppenbeck2013, Wienholtz2013, Michimasa2018}. Precise spectroscopic measurements of $\mu_I$ are again critical, as typically one uses a reference nucleus of the same element to deduce unknown properties of other isotopes \cite{Klose2019}. 

The interaction of an atom with electronic angular momentum $\mathbf{J}$ and nuclear spin $\mathbf{I}$, and a static magnetic field $\mathbf{B}$ is described to good approximation by the Hamiltonian \cite{Corney1978}  
\begin{eqnarray} \label{eq:Hamiltonian}
H &=& hA\mathbf{I}\cdot\mathbf{J} - \left(\mathbf{\boldsymbol{\mu}_J} + \mathbf{\boldsymbol{\mu}_I}\right)\cdot\mathbf{B}~, \nonumber \\
&=& hA\mathbf{I}\cdot\mathbf{J} + \left(g_J\mu_{\rm{B}}\mathbf{J} - g_I\mu_{\rm{N}}\mathbf{I}\right)\cdot\mathbf{B}~,
\end{eqnarray}
where $h$ is Planck's constant, $A$ is the magnetic dipole hyperfine interaction constant, $g_J$ and $g_I$ are the electronic and nuclear $g$-factors, and $\mu_{\rm{B}}$ and $\mu_{\rm{N}}$ are the Bohr and nuclear magnetons. Note that the apparent sign change arises from the conventional definitions of $\mathbf{\boldsymbol{\mu}_J}$ and $\mathbf{\boldsymbol{\mu}_I}$. The first term describes the magnetic dipole interaction between the nucleus and bound electrons, and the second and third terms describe the Zeeman interaction between the static magnetic field and the electronic and nuclear magnetic moments respectively. The energy eigenstates of this Hamiltonian, in general, must be numerically calculated. However, for the case where $J=1/2$, the eigenstate energies are given analytically by the Breit-Rabi formula \cite{Breit1931, Corney1978}
\begin{eqnarray}\label{eq:BR}
E_{\pm}\left(B,M\right) = &-&\frac{E_{\rm{hfs}}}{2(2I+1)} - g_I\mu_{\rm{N}}BM \nonumber \\
&\pm& \frac{E_{\rm{hfs}}}{2}\sqrt{1+\frac{4\chi BM}{2I+1}+\chi^2B^2}~, 
\end{eqnarray}  
where $M = M_I + M_J$ is the magnetic quantum number, $E_{\rm{hfs}} = hA\left(I+1/2\right)$ is the zero-field energy splitting between the two hyperfine manifolds, and ${\chi = \left(g_I\mu_{\rm{N}} + g_J\mu_{\rm{B}}\right)/E_{\rm{hfs}}}$. The Breit-Rabi formula highlights that the transition frequency between states in the $J=1/2$ manifold depends upon three intrinsic properties of the atomic ground level; namely the zero-field hyperfine splitting, and the nuclear and electronic magnetic moments. Spectroscopy of ground-level hyperfine structure is therefore an excellent technique to test physical theories of atomic constituents. 

Many spectroscopic techniques have been used to determine these constants. The earliest techniques used spectroscopy in thermal atomic beams (see Arimondo \textit{et al.} \cite{Arimondo1977} for a detailed review), or nuclear-magnetic-resonance (NMR) spectroscopy. The accuracy of NMR measurements is limited by the systematic error caused by the `chemical shift' \cite{Pople1957I, Pople1957II, Kaupp2004}, which describes the magnetic shielding of the target nucleus by the solution in which the atom is measured. The chemical shift is challenging to calculate \cite{Kaupp2004} and has led to significant disagreements between theory and experiment \cite{Skripnikov2018}. The advent of ion trapping extended the possibility of measurements to ions, consequently enhancing the achievable precision due to the ability to confine ions to a small region of free space which minimises effects such as magnetic field inhomogeneities, removes chemical shifts, and provides extremely long coherence times \cite{Werth1995, Savard2000, Karr2009}. 

$^{43}$Ca is the only naturally occurring calcium isotope with non-zero nuclear spin ($I=7/2$), making it an ideal reference nucleus. All three of the relevant atomic constants have been previously measured. Arbes \textit{et al.} \cite{Arbes1992} measured ${E_{\rm{hfs}} = \SI{3225608286.4 \pm 0.3}{Hz}}$ using double-resonance spectroscopy of $^{43}$Ca$^+$ ions in a Paul trap, and Tommaseo \textit{et al.} \cite{Tommaseo2003} measured ${g_J = \num{2.00225664 \pm 0.00000009}}$ using double-resonance spectroscopy of $^{40}$Ca$^+$ ions in a Penning trap. One would expect the isotopic dependence of $g_J$ to be smaller than the experimental measurement uncertainty, based upon similar measurements using Ba$^+$ isotopes \cite{Marx1998}. There exist three previous measurements of $\mu_I$ for $^{43}$Ca which are summarised in table \ref{tab:gI}. Two measurements were made using NMR of liquid Ca salts \cite{Jeffries1953, Lutz1973}, and the other by spectroscopy in an atomic vapour \cite{Olschewski1972}.   
  
\begin{table}
\begin{ruledtabular}
\begin{tabular}{d{2.8}cc}
\multicolumn{1}{c}{$\mathbf{\boldsymbol{\mu}_I/\boldsymbol{\mu}_\mathbf{N}}$} & \textbf{Environment} & \textbf{Reference} \\
\colrule
\rule{0pt}{2.2ex}
-1.3152(2) & Liquid NMR, Ca$^{2+}$ & \cite{Jeffries1953} \\
-1.315645(7) & Liquid NMR, Ca$^{2+}$ & \cite{Lutz1973} \\
-1.31537(60) & Atomic vapour, Ca & \cite{Olschewski1972}\\ 
-1.315350(9)(1) & Single trapped ion, Ca$^{+}$ & This work\\ 
-1.315349(9)(4) & Single trapped ion, Ca$^{+}$ & This work\\ 
\end{tabular}
\end{ruledtabular}
\caption{\label{tab:gI} Measurements of the nuclear magnetic moment of $^{43}$Ca in units of $\mu_{\rm{N}}$, and the environment in which they were measured. Note that all of these measurements are uncorrected for diamagnetic shielding due to bound electrons, and the NMR measurements are uncorrected for chemical shifts. With respect to the results obtained in this work, the two uncertainties correspond to the contribution due to the uncertainty in the ground-level hyperfine splitting $E_{\rm{hfs}}$ measured by Arbes \textit{et al}. \cite{Arbes1992}, and the uncertainty in our measurements respectively.}
\end{table} 

This paper details two precision measurements of $\mu_I$ using a single $^{43}{\rm{Ca}}^{+}$ ion held in a surface-electrode Paul trap with integrated microwave circuity \cite{Allcock2013}. Ramsey spectroscopy \cite{Ramsey1950} was performed on the $\left|F=4, M=0\right\rangle \rightarrow \left|F=3, M=1\right\rangle$ clock transition \cite{Langer2005}, at ${B = \SI{146.094}{G}}$, where the coherence time is of the order of minutes \cite{Harty2014} due to the lack of first-order magnetic field sensitivity. The two measurements were taken nine months apart in the same apparatus, and used either the time or frequency variants of Ramsey spectroscopy. The remainder of this paper is structured as follows. We initially describe the experimental apparatus and measurement techniques used, before analysing in detail sources of systematic uncertainty in our measurement. We then report the systematic-corrected measurement of $\mu_I$, and compare to previous measurements. We estimate the diamagnetic-corrected nuclear moment using published calculations of the shielding constants for Ca \cite{Fuller1976} and Ca$^{2+}$ \cite{Antuvsek2013}. From our measurement, and previous NMR spectroscopy \cite{Lutz1973}, we extract the change in shielding constant of between a free Ca$^+$ ion and Ca$^{2+}$ ions in D$_2$O solution.

\begin{figure}
\includegraphics{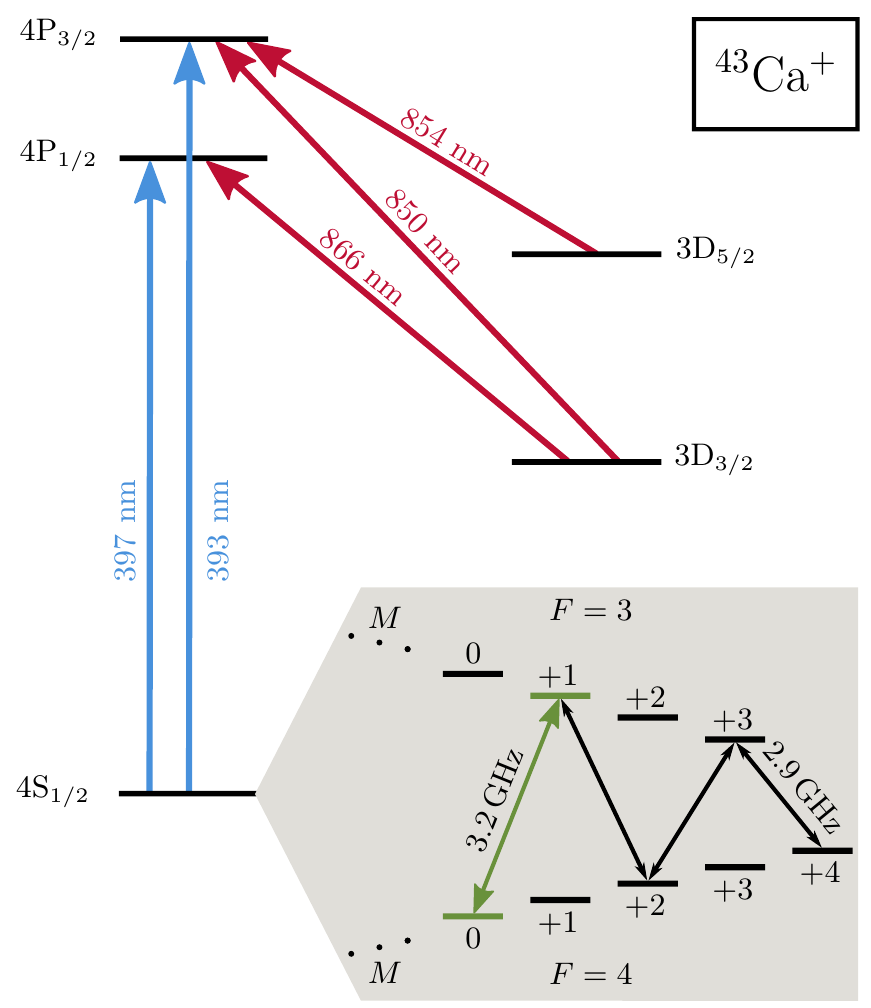}
\caption{\label{fig:energy_levels}Energy level diagram of $^{43}{\rm{Ca}}^+$ at $B= \SI{146.094}{G}$ showing the relevant states for laser cooling, state preparation, and readout. The lower panel shows the hyperfine structure of the 4S$_{1/2}$ state. The green states show the clock transition used for the measurements in this work, and the thin black arrows show the transitions used for state preparation and readout.}
\end{figure}

A single $^{43}{\rm{Ca}}^+$ ion was loaded into a surface trap \cite{Allcock2013} from a $\SI{12}{\percent}$ isotopically enriched calcium source using isotope-selective photoionization \cite{Lucas2004}. The ion was Doppler cooled on the $\SI{397}{nm}$ $4{\rm{S}}_{1/2} \rightarrow 4{\rm{P}}_{1/2}$ transition, and repumped by an $\SI{866}{nm}$ laser to close the cooling cycle (see figure \ref{fig:energy_levels}). The RF trapping field was driven at $\SI{38.7}{MHz}$, and the radial secular frequency was varied between $\SI{2}{MHz}$ and $\SI{4}{MHz}$. The axial frequency was $\SI{0.5}{MHz}$.  A static magnetic field was provided by current flowing through two coils external to the vacuum chamber. Transitions between energy levels in the ground-state hyperfine manifolds were driven directly by the magnetic field generated from a current applied to one of the trap's integrated microwave electrodes. The microwaves were synthesized using a commercial microwave synthesizer phase-locked to a rubidium frequency standard (RbFS \footnote{Stanford Research Systems FS725}). To ensure an accurate absolute frequency measurement, the phase difference between the RbFS and a GPS-disciplined oscillator (GPSDO \footnote{Trimble Thunderbolt GPS-disciplined oscillator}) was measured for a period of $\SI{12}{hrs}$ directly after the clock transition frequency measurements. At this time interval, the GPSDO has a frequency stability of $\num{1.16e-12}$. 

Additional lasers were used for state preparation and readout \cite{Myerson2008, Harty2014}. We first optically pumped the ion to the $\left|F=4, M=4\right\rangle$ state using circularly-polarized $\SI{397}{nm}$ light, after which the ion was prepared into the $\left|F=3, M=1\right\rangle$ state via a series of microwave $\pi$ pulses (see figure \ref{fig:energy_levels}). After experiments on the clock transition, microwave $\pi$ pulses transferred population in the $\left|F=3, M=1\right\rangle$ state back to the $\left|F=4, M=4\right\rangle$ state. Population in the $\left|F=4, M=4\right\rangle$ state was then shelved in the metastable 3D$_{5/2}$ level by a series of $\SI{393}{nm}$ and $\SI{850}{nm}$ pulses (see \cite{Myerson2008} for further details). The Doppler cooling beams where then applied and the state of the ion inferred by the absence or presence of ion fluorescence. The ion was repumped to the ground level using an $\SI{854}{nm}$ pulse.     

The applied static magnetic field was measured using Rabi spectroscopy on the stretch $\left|F=4, M=4\right\rangle \rightarrow \left|F=3, M=3\right\rangle$ transition. This transition is first-order sensitive to the applied magnetic field, with a transition frequency sensitivity of ${df/dB = \SI{-2.36}{MHz/G}}$. After optical pumping into the $\left|F=4, M=4\right\rangle$ state, a microwave $\pi$-pulse was applied and the probability of remaining in the $\left|F=4, M=4\right\rangle$ state was measured using electron shelving \cite{Myerson2008, Harty2014}. The pulse $\pi$-time was $\SI{22.2}{\micro s}$, leading to a transition FWHM of $\SI{36}{kHz}$. The coil current was then adjusted to ensure agreement between the measured stretch transition frequency and that predicted by the Breit-Rabi formula at the clock field of $\SI{146.094}{G}$. The experimental uncertainties in $E_{\rm{hfs}}$, $g_I$, and $g_J$ contribute a systematic frequency shift of the stretch transition which is three orders of magnitude smaller than the systematic shift induced by magnetic field fluctuations. This procedure enabled calibration of the static magnetic field to an accuracy of $\approx \SI{1}{mG}$. 

\begin{figure*}
\includegraphics[width=17.8cm]{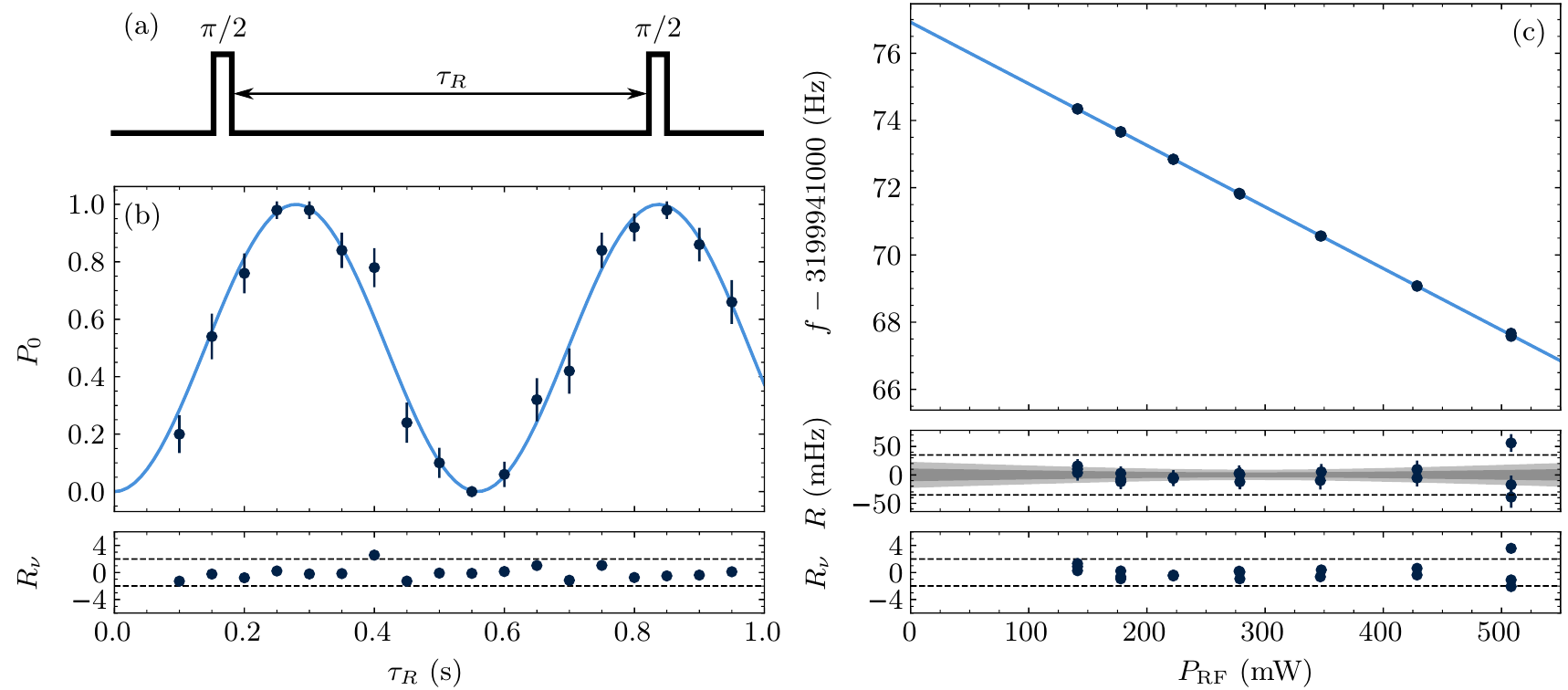}
\caption{\label{fig:Ramsey_D3}(a) Pulse sequence used for Ramsey interferometry. The ion is first prepared in the $\left|F=3, M=1\right\rangle$ state before applying a pair of $\pi /2$-pulses of duration $\SI{20.9}{\micro s}$, at a frequency $(f_{\upmu \rm{W}})$ and separated by a Ramsey delay $\tau_R$. The ion's resulting state was subsequently measured using electron shelving. Figure (b) shows a typical Ramsey spectroscopy signal on the clock transition at a trap RF power ($P_{\rm{RF}}$) of $\SI{279}{mW}$, driven with a $\pi/2$-pulse duration of $\SI{20.9}{\micro s}$ and a microwave frequency of $f_{\upmu \rm{W}} = \SI{3199941070}{Hz}$. The error bars are from quantum projection noise. The solid line is a fit to the data using an analytical expression for the propagator of a Ramsey interferometry sequence, where the free parameters are the frequency offset, $\SI{-1.791(12)}{Hz}$, and an amplitude scale factor. Figure (c) shows the measured clock transition frequency $f$ as a function of applied trap RF power. The solid line is a weighted straight-line fit to the data. The lower figures show the residuals $R$, and the normalised residuals $R_{\nu}$. Error bars represent the combined uncertainty from the Ramsey fit and from the digitisation error of the RF power meter. The smaller and larger shaded regions in the plot of residuals shows the $\SI{68}{\percent}$ and $\SI{95}{\percent}$ confidence intervals of the linear fit respectively, the dashed lines illustrate the $\pm\SI{35}{mHz}$ total systematic uncertainty, and the dashed lines in the plot of the normalised residuals show $R_{\nu} = \pm 2$. We calculate the reduced chi-squared statistic to be $\chi^2_{\nu} = 0.9$ (b) and $\chi^2_{\nu} = 1.5$ (c). Note that the measured fractional frequency deviation between the RbFS and the GPSDO of $f_{\rm{RbFS}}/f_{\rm{GPSDO}} = \num{3.22\pm0.02 e-10}$, corresponding to a systematic shift of $\SI{1.030 \pm 0.008}{Hz}$ of the clock transition frequency, has been corrected in the data.}
\end{figure*}

Ramsey spectroscopy \cite{Ramsey1950} (see figure \ref{fig:Ramsey_D3}(a)) was used to measure the clock transition frequency $(f)$, as the long coherence time facilitates large Ramsey delays ($\tau_R$) and hence enhanced precision. The two measurements of the clock transition frequency used differing variants of Ramsey spectroscopy. The first measurement consisted of Ramsey interferometry using a fixed microwave frequency $(f_{\upmu \rm{W}})$ and a variable Ramsey delay, whilst the second measurement consisted of varying the microwave frequency with a fixed Ramsey delay. Using two variants of Ramsey spectroscopy gives us increased confidence in our evaluation of systematic errors. 

The first measurement of the clock qubit transition frequency was performed as follows. The magnetic field was set to $\SI{146.094}{G}$ using the stretch transition $\left|F=4, M=4\right\rangle \rightarrow \left|F=3, M=3\right\rangle$ spectroscopy method outlined previously. The ion was then prepared in the $\left|F=3, M=1\right\rangle$ state via a series of microwave transfer pulses (see figure \ref{fig:energy_levels}). A pair of $\pi/2$-pulses was subsequently applied at a fixed frequency of $f_{\upmu \rm{W}}$ near $\SI{3.199941070}{GHz}$, with a variable Ramsey delay $\tau_R = \SI{0.1}{s} \rightarrow \SI{1}{s}$. We inferred a fractional frequency deviation between the RbFS and the GPSDO of $f_{\rm{RbFS}}/f_{\rm{GPSDO}} = \num{3.22\pm0.02 e-10}$, corresponding to a systematic shift of $\SI{1.030 \pm 0.008}{Hz}$ of the measured clock transition frequency. A typical Ramsey spectroscopy signal of the clock transition is shown in figure \ref{fig:Ramsey_D3}(b). The solid line is a fit to the data using an analytical expression for the propagator of a Ramsey interferometry sequence, where the only free parameters were the frequency offset and an amplitude scale factor to account for imperfect state readout. The lower figures show a plot of the normalised residuals $R_{\nu}$ \cite{Hughes2010}, defined as the residuals between the data and the fit, normalised to their respective uncertainty. 

Figure \ref{fig:Ramsey_D3}(c) shows the measured clock transition frequency as a function of applied trapping RF power ($P_{\rm{RF}}$). Currents flowing in the RF trapping electrodes generate oscillating magnetic fields at the RF drive frequency $\left(\SI{38.7}{MHz}\right)$. These oscillating magnetic fields off-resonantly couple states within each hyperfine manifold, which are separated by approximately $\SI{50}{MHz}$, resulting in a systematic shift of the clock transition frequency. The unperturbed transition frequency in the absence of the AC Zeeman shift caused by the trapping RF is therefore determined from the intercept of the straight-line fit in figure \ref{fig:Ramsey_D3}(c). The figure shows the frequency shift is of the order of several Hz. This is larger than one might expect in comparison to an ion frequency standard \cite{Berkeland1998}, as ion frequency standards typically use a Paul trap where the RF electrodes are symmetric about the ion and therefore the generated magnetic fields are nulled at the ion. However, in a surface trap the null is in the plane of the trap and therefore the magnitude of cancellation is greatly reduced. The residuals show excellent agreement between theory and experiment; however they also highlight an outlier point at $P_{\rm{RF}}\approx \SI{500}{mW}$. This point lies above the fitted line by $\approx \SI{50}{mHz}$, and may indicate drifts in systematic shifts over the timescale of a full data collection cycle; it is consistent with our systematic uncertainty (see below).

\begin{figure}
\includegraphics[width=\columnwidth]{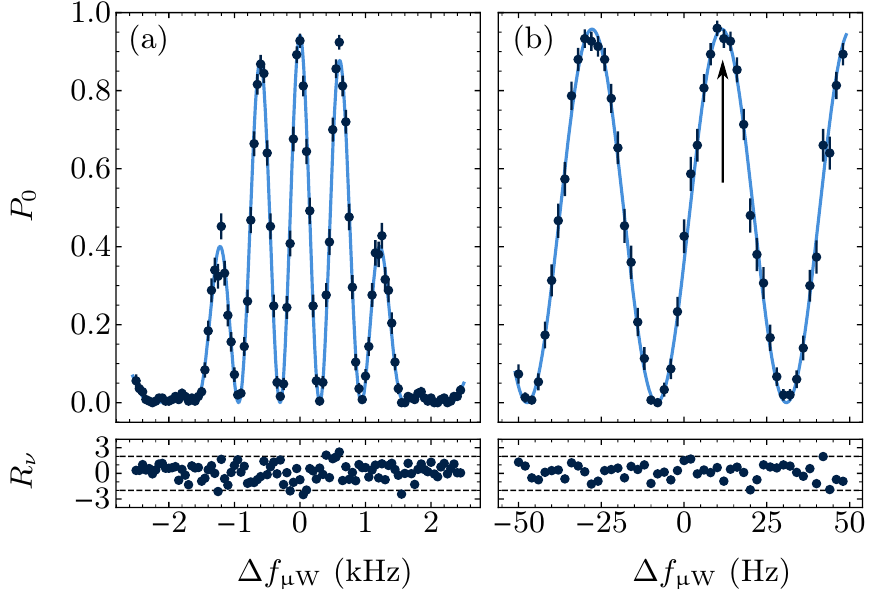}
\caption{\label{fig:Ramsey_TPH} Typical Ramsey spectroscopy signals on the clock transition using $\tau_R = \SI{1.05}{ms}$ (a) and $\tau_R = \SI{24.93}{ms}$ (b), a $\pi/2$-pulse duration of $\SI{478.62}{\micro s}$, a microwave frequency $f_{\upmu \rm{W}} = \SI{3199941071}{Hz}$, and static magnetic field offset from the clock field $\Delta B = \SI{100}{mG}$. The error bars are from quantum projection noise. The vertical arrow in (b) indicates the central fringe. The solid line is a fit to the data using an analytical expression for the propagator of a Ramsey interferometry sequence, where the free parameters are the frequency offset and an amplitude scale factor. The lower figures show a plot of the normalised residuals $R_{\nu}$. We calculate the reduced chi-squared statistic to be $\chi^2_{\nu} = 1.04$ (a) and $\chi^2_{\nu} = 0.84$ (b), and the dashed lines show $R_{\nu} = \pm 2$.}
\end{figure}

The second measurement of the clock transition frequency was performed nine months later in the same apparatus. The clock transition frequency was measured as a function of applied magnetic field about $\SI{146.094}{G}$, using a fixed trap RF power. The ion was prepared in the $\left|F=3, M=1\right\rangle$ state, after which a pair of $\pi/2$-pulses was applied at a variable detuning $\Delta f_{\upmu \rm{W}}$ about the clock transition frequency with a fixed Ramsey delay. Figure \ref{fig:Ramsey_TPH} shows a typical Ramsey spectroscopy signal of the clock transition at a static magnetic field offset from the clock field $\Delta B = \SI{100}{mG}$ using Ramsey delays $\tau_R = \SI{1.05}{ms}$ (a) and $\tau_R = \SI{24.93}{ms}$ (b), identifying a frequency shift of $\SI{11.34 \pm 0.08}{Hz}$. The short Ramsey delay was used to identify the central fringe, and the longer delay was used for the precision measurements. The solid line is once again a fit to the data using an analytical expression for the propagator of a Ramsey interferometry sequence, where the only free parameters were the frequency offset and an amplitude scale factor to account for imperfect state readout.

\begin{figure}
\includegraphics[width=\columnwidth]{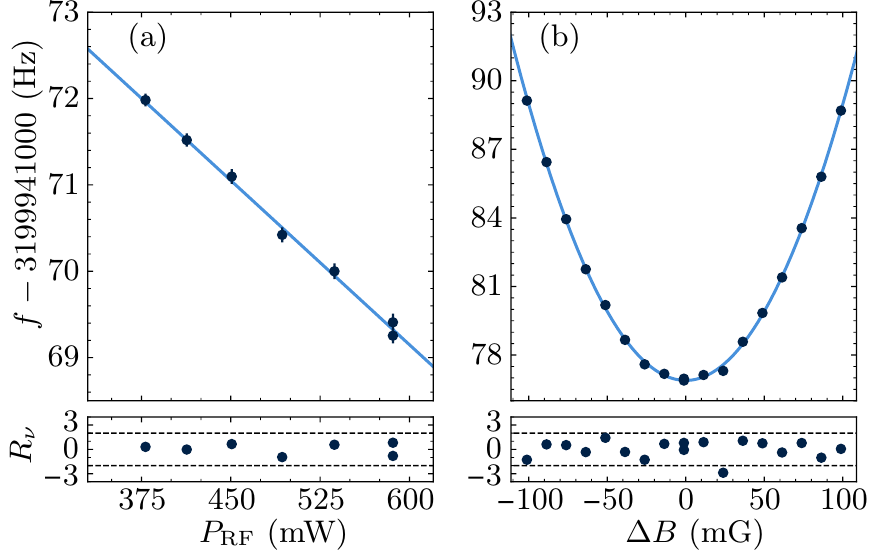}
\caption{\label{fig:BR_fit}(a) Clock-transition frequency shift as a function of applied trap RF power, where the solid line is a linear fit to the data. (b) Measured clock transition frequency as a function of applied static magnetic field offset from $\SI{146.094}{G}$, measured at $P_{\rm{RF}}= \SI{392}{mW}$. The systematic shift due to the RF induced AC Zeeman shift ($\SI{-5.050 \pm .12}{Hz}$) and the frequency deviation between the RbFS and the GPSDO ($\SI{1.322 \pm 0.008}{Hz}$) have been applied to all points equally. The solid line is a fit to the experimental data using the Breit-Rabi formula with only $g_I$ as a fitted parameter. The lower figures show the normalised residuals, from which we calculate $\chi^2_{\nu} = 0.8$ (a) and $\chi^2_{\nu} = 1.3$ (b) respectively. The dashed lines show $R_{\nu} = \pm 2$.}
\end{figure}

To determine the AC Zeeman shift of the clock transition frequency, we measured the clock transition frequency using Ramsey spectroscopy for a series of trap RF powers, as shown in figure \ref{fig:BR_fit}(a). We observed a linear frequency shift as a function of trap RF power, corresponding to a systematic shift of $\SI{-5.050 \pm .12}{Hz}$ at the RF power used during the subsequent experiments. We also measured a fractional frequency deviation between the RbFS and the GPSDO of $f_{\rm{RbFS}}/f_{\rm{GPSDO}} = \num{4.13\pm0.02 e-10}$, corresponding to a systematic shift of $\SI{1.322 \pm 0.008}{Hz}$ of the clock transition frequency. Figure \ref{fig:BR_fit}(b) shows the measured clock transition frequency as a function of static magnetic field about $\SI{146.094}{G}$, corrected for the systematic shifts discussed previously. The solid line is a fit to the data using the Breit-Rabi formula from which we are able to extract the clock transition frequency and hence the nuclear magnetic moment. The only fit parameter was $g_I$, and values of $E_{\rm{hfs}}$ and $g_J$ were taken from Arbes \textit{et al.} \cite{Arbes1992} and Tommaseo \textit{et al.} \cite{Tommaseo2003} respectively. The lower figure shows a plot of the normalised residuals, from which we calculate $\chi^2_{\nu} = 1.3$, indicating an excellent agreement between theory and experiment. 
 
The RF-induced AC Zeeman shift and the calibration of the RbFS were the major sources of systematic frequency shifts in these measurements. We now discuss the uncertainties in these shifts, and other sources of uncertainty (see table~\ref{tab:errors}). 

The experimental uncertainty in the comparison of the RbFS with the GPSDO, together with the specified instability of the GPSDO, give a total uncertainty of $\SI{8}{mHz}$ in the RbFS calibration.

The largest uncertainty in correcting for the RF-induced AC Zeeman shift arises from any potential non-linearity in the RF power meter (Keysight V3500A) which was used to measure the applied trap RF power. By comparing the device to a precision power sensor (Keysight N8481A), we measure any non-linearity to be $<\SI{0.015}{dB}$ over the range used in the experiment. We bound the uncertainty by re-fitting the data in figures~\ref{fig:Ramsey_D3}(c) using a worst-case scenario in which the gradient is maximally affected. We find an upper limit on the uncertainty of the fitted intercept of $\pm \SI{25}{mHz}$. The power meter also has a $\pm \SI{0.005}{dB}$ digitization error which, from the slope in figure 2(c), effectively gives a random frequency error between $\pm \SI{3}{mHz}$ and $\pm \SI{11}{mHz}$ for the range of powers used; this has been accounted for in the error bars in figure 2(c).

Drifts in RF power in the trap, relative to the measured input power, can occur for example due to variation in the step-up of the RF resonator. We can estimate such drifts by monitoring the ion's radial secular frequency, as $\omega_r\propto\sqrt{P_{\rm{RF}}}$. We measured short-term variations in $P_{\rm{RF}}$ at the $\SI{0.05}{\percent}$ level, and longer-term drift of $\SI{0.05}{\percent}$/hour~\cite{Allcock2011}. As the data in figure~\ref{fig:Ramsey_D3}(c) were taken mostly in time order, over $\approx 5$~hours, a linear drift in time could systematically affect the fitted gradient. We determine the worst-case effect to be $\pm \SI{10}{mHz}$.

The RF-induced AC Zeeman shift is dependent upon the magnitude and polarization of the RF magnetic field at the ion position. Hence drifts of the ion's radial position during the experiments can lead to changes in the measured shift. To quantify this effect, we measured the clock transition frequency as a function of radial DC compensation fields $E_x$ and $E_y$. We observe frequency shifts of $df/dE_x = \SI{3.9 \pm 0.2}{mHz/(V/m)}$ and $df/dE_y = \SI{2.34 \pm 0.07}{mHz/(V/m)}$ in the directions parallel and perpendicular to the trap surface respectively \footnote{The gradients $df/dE$ are independent of trap RF power $P_{\rm{RF}}$: we measure the AC Zeeman shift to vary linearly for small displacements $x$ of the ion, so that it is proportional to $xP_{\rm{RF}}$, while $x\propto 1/P_{\rm{RF}}$ for a given $E$.}. Experimentally, we observed that the micromotion compensation was stable to within $\pm \SI{5}{V/m}$ and therefore assign an upper bound of $\pm \SI{20}{mHz}$ to this systematic shift. 

For larger excursions in ion position, such as could occur due to micromotion or thermal motion of the ion, we measure a quadratic dependence of the AC Zeeman shift on position~\cite{Harty2013}, which will not average to zero. We observed intrinsic (uncompensatable) micromotion in the $y$ direction of amplitude $\approx \SI{20}{nm}$, which leads to negligible shift. Based on the measured heating rate of the trap~\cite{Sepiol2016}, the radial temperature of the ion is $T<\SI{10}{mK}$ after the longest ($\SI{1}{s}$) Ramsey delays, which would lead to a shift $<\SI{1}{mHz}$. 

\begin{table*}
\begin{ruledtabular}
\begin{tabular}{cccc}
\rule{0pt}{2.2ex} \rule[-1.2ex]{0pt}{0pt}\textbf{Measurement} & \textbf{Source} & \textbf{Magnitude (mHz)} & \textbf{Uncertainty (mHz)} \\
\colrule
\rule{0pt}{2.2ex}
\hspace{-0.7ex}1 & Rb frequency standard & 1030 & 8 \\
\hline
2 & Rb frequency standard & 1322 & 8 \\
2 & RF-induced AC Zeeman & 5050 & 120 \\
\hline
1, 2 & RF power meter linearity & 0 & 25 \\
1, 2 & RF power drift & 0 & 10 \\
1, 2 & Ion position drift & 0 & 20 \\
1, 2 & Ion position (thermal motion) & $<1$ & 1 \\
1, 2 & Magnetic field ($\SI{50}{Hz}$) & 11 & 8 \\
1, 2 & Magnetic field (DC error) & 1 & 1 \\
1, 2 & Magnetic field (non-$\SI{50}{Hz}$) & $<0.1$ & 0.1 \\
1, 2 & RF-induced AC Stark & $< 0.01$ & 0.01 \\
1, 2 & Blackbody AC Stark & $<0.01$ & 0.01 \\
1, 2 & Second-order Doppler & $<0.001$ & 0.001\\ 
\end{tabular}
\end{ruledtabular}
\caption{\label{tab:errors}Systematic frequency shifts of the measured clock transition frequency. The measurement column indicates to which measurement the shift applies, where the numbering follows the order of measurements presented in the main text. Note that for measurement 1, we extrapolate to zero RF-induced AC Zeeman shift; systematic uncertainty introduced by this extrapolation is accounted for by the values in the Table.}
\end{table*} 

Fluctuations in the static magnetic field $B$ can cause frequency shifts in the measured clock transition frequency $f$. These enter at second order for the clock transition, for which $d^2f/dB^2 = \SI{2.416}{mHz/mG^2}$. The automated servo corrections to the static field which were applied during the experiment in figure~\ref{fig:Ramsey_D3} had an rms amplitude of $\SI{0.8}{mG}$, implying a frequency shift of $\approx \SI{1}{mHz}$ due to servo imprecision. A larger source of magnetic field noise is caused by $\SI{50}{Hz}$ mains power. We measured the effect of $\SI{50}{Hz}$ field modulation by performing Ramsey spectroscopy on the first-order sensitive $\left|F=4, M=4\right\rangle \rightarrow \left|F=3, M=3\right\rangle$ transition as a function of delay after the zero-crossing of the mains power cycle (line trigger). The measured transition frequency thus reveals the change in local magnetic field within each mains cycle. We measured an rms amplitude of $\SI{3 \pm 1}{mG}$, which contributes a systematic shift of $\SI{11 \pm 8}{mHz}$ to the clock transition frequency. Field fluctuations incoherent with $\SI{50}{Hz}$ were bounded by comparing the measured spectral width of the stretch transition (whilst line-triggering to remove $\SI{50}{Hz}$ effects) to the theoretical width for our measured Rabi frequency. We were not able to measure a difference in spectral width within the experimental uncertainty of $\SI{0.6}{kHz}$, which bounds shifts due to such field fluctuations to be $< \SI{0.1}{mHz}$.

Residual electric fields at the ion can induce an AC Stark shift, of approximate magnitude $\SI{1e-11}{Hz/(V/m)^2}$ \cite{Itano1982}. The intrinsic micromotion observed in the $y$ direction implies an RF field of rms amplitude $\SI{400}{V/m}$ \footnote{The radial thermal motion considered above will only slightly increase the rms RF field experienced by the ion}. Hence the AC Stark shift due to trap fields is expected to be $<\SI{10}{\micro Hz}$. The rms electric field due to black body radiation at $\SI{300}{K}$ is $\SI{830}{V/m}$ \cite{Itano1982}, leading to a similar shift. 

A second-order Doppler shift will also be present, due to the intrinsic micromotion and any thermal motion of the ion. The total shift is $<\SI{1}{\micro Hz}$. 

We add the various systematic uncertainties in quadrature to obtain a total systematic uncertainty, which we add linearly to the statistical uncertainty for each measurement, to report both the clock transition frequency and the nuclear magnetic moment. We extract the clock transition frequency from the first measurement by determining the intercept in figure \ref{fig:Ramsey_D3}(c). This gives $f = \SI[parse-numbers=false]{3199941076.920 \pm 0.011 _{stat} \pm 0.035 _{syst}}{Hz}$, which corresponds to a nuclear magnetic moment of $\mu_I / \mu_{\rm{N}} = -1.315350(9)(1)$. We extract the clock transition frequency and the nuclear magnetic moment from the second measurement by fitting the Breit-Rabi formula to the data presented in figure \ref{fig:BR_fit}(b), where the only fit parameter was $g_I$. This is determined to be $f = \SI[parse-numbers=false]{3199941076.89 \pm 0.02 _{stat} \pm 0.13_{syst}}{Hz}$, which corresponds to a nuclear magnetic moment of $\mu_I / \mu_{\rm{N}} = -1.315349(9)(4)$. The two uncertainties on the nuclear magnetic moment are due to the uncertainty in the ground-level hyperfine splitting $E_{\rm{hfs}}$ measured by Arbes \textit{et al}., and the uncertainty in our measurements respectively. Note that these measurements are not corrected for diamagnetic shielding of the nucleus due to the bound electrons \cite{Lamb1941, Pyykko2000}. The two measurements are in excellent agreement with each other which gives confidence in the reported values given the differing spectroscopic methods and the nine month interval between measurements. 

The measured nuclear magnetic moment $\mu_{\rm{meas}}$ differs from that of the bare nucleus $\mu_{\rm{bare}}$ because of diamagnetic shielding by the bound electrons. A calculation of the diamagnetic correction factor, $\mu_{\rm{bare}}/\mu_{\rm{meas}}$, for Ca$^+$ has not been published. However, there are published values for both Ca \cite{Fuller1976} and Ca$^{2+}$ \cite{Antuvsek2013}. We approximate the correction for Ca$^+$ to be the mean of these values, and assume a $\SI{95}{\percent}$ confidence interval between the Ca (1.001495) and Ca$^{2+}$ (1.001465) correction factors, resulting in a diamagnetic correction of $\num{1.001480(8)}$ for Ca$^{+}$. This is consistent with a recent unpublished calculation \cite{AntusekPersComm2021}. Applying this correction to our measurement, we calculate a corrected magnetic moment of $\mu_{\rm{bare}} / \mu_{\rm{N}} = -1.317297(13)$.

Our measurement (before diamagnetic correction) is consistent with previous measurements by Jeffries \textit{et al.} \cite{Jeffries1953} and Olschewski \textit{et al.} \cite{Olschewski1972}, but disagrees with the precision NMR measurement by Lutz \textit{et al.} \cite{Lutz1973} (see Table \ref{tab:gI}). The measurement by Lutz \textit{et al.} is, like our measurement, uncorrected for diamagnetic shielding due to the bound electrons. However, it is also uncorrected for the change in shielding constant due to the D$_2$O solution in which the Ca$^{2+}$ ions were measured, to which we attribute the discrepancy \footnote{We note that the differences in the diamagnetic shielding constants between Ca$^+$ and Ca$^{2+}$ given above are an order of magnitude too small to explain the discrepancy.}. Lutz \textit{et al.} used the free-atom measurement of Olschewski \textit{et al.} \cite{Olschewski1972} to determine the shielding constant, defined as $\sigma = 1 - (\mu_{\rm{meas}}/\mu_{\rm{bare}})$. However, the measurement by Olschewski \textit{et al.} \cite{Olschewski1972} does not have a diamagnetic correction applied, and therefore Lutz \textit{et al.} were in fact measuring the change in shielding constant between a free atom and a solvated ion. We define this as $\Delta \sigma = \sigma_{\rm{NMR}} - \sigma_{\rm{atom}}$ where the subscripts refer to the measurement technique. Due to the large uncertainty in $\mu_{\rm{atom}}$, Lutz \textit{et al.} reported a measurement of $\Delta \sigma = \num{-0.0002 \pm 0.0005}$ which is consistent with zero. With our precision measurement of $\mu_I$ we can evaluate the change in shielding constant between a free Ca$^+$ ion and a solvated Ca$^{2+}$ ion to be ${\Delta \sigma = \num{-0.00022 \pm 0.00001}}$. By comparing the calculated shielding constant of Ca$^{2+}$ in H$_2$O \cite{Antuvsek2013} and that estimated above for Ca$^+$, we are able to extract a theoretical change of shielding constant of $\Delta \sigma = \num{-0.00022 \pm 0.00004}$, which is consistent with our measurements.

The precision of our nuclear magnetic moment measurement is limited by the uncertainty in the ground-level hyperfine splitting measured by Arbes \textit{et al}. This measurement could be improved in the following way. The sensitivity of a hyperfine transition to $\mu_I$ is dominated by the second term of equation \ref{eq:BR}. Therefore, by applying the demonstrated techniques to a $\Delta M = 0$ transition, such as the $\left|F=4, M=1\right\rangle \rightarrow \left|F=3, M=1\right\rangle$ clock transition at ${B = \SI{287.783}{G}}$, it would be possible to improve upon the current measurement of $E_{\rm{hfs}}$ and hence the overall precision of $\mu_I$. To increase precision further, the size of the RF-induced AC Zeeman shift could be reduced by performing the measurements in a 3D Paul trap with a symmetric trap design to minimize RF magnetic fields at the ion. With increased precision, it may be necessary to take into account modifications to the Breit-Rabi formula \cite{Shiga2011}.

We note that these measurement techniques could also be applied to more exotic calcium isotopes with nuclear spin. For example, $^{41}$Ca \cite{Hasegawa2006, Kitaoka2013} and $^{45}$Ca can both be readily artificially produced, and have long enough half-lives to be used in similar experiments.

In summary, we have performed two precision measurements of the nuclear magnetic moment of a single \textsuperscript{43}Ca\textsuperscript{+} trapped in a surface-electrode Paul trap. These measurements improve upon previous free-atom experiments \cite{Olschewski1972} by more than one order-of-magnitude, and are free from systematic shifts associated with NMR measurements \cite{Jeffries1953, Lutz1973}. This measurement adds to the increasing number of precision nuclear moment measurements, and improves the accuracy of measurements in the calcium isotopic chain which are critical in tests of fundamental physics theories.

We thank Derek Stacey and Andrej Antu\v{s}ek for helpful discussions. This work was supported by the U.S.\ Army Research Office (contract no. W911NF-18-1-0340) and UK EPSRC.

\bibliographystyle{unsrt}
\bibliography{bibfile}

\end{document}